# Magnetic-Related States and Order Parameter Induced in a Conventional Superconductor by Nonmagnetic Chiral Molecules

**Hen Alpern,[1,2] Konstantin Yavilberg,[3] Tom Dvir,[1] Nir Sukenik,[2] Maya Klang,[1] Shira Yochelis,[2] Hagai Cohen,[4] Eytan Grosfeld,[3] Hadar Steinberg,[1] Yossi Paltiel[2] and Oded Millo[1]**

[1] Racah Institute of Physics and the Center for Nanoscience and Nanotechnology, The Hebrew University of Jerusalem, Jerusalem 91904, Israel

[2] Applied Physics Department the Center for Nanoscience and Nanotechnology, The Hebrew University of Jerusalem, Jerusalem 91904, Israel

[3] Department of Physics, Ben-Gurion University of the Negev, Beer-Sheva 84105, Israel

[4] Chemical Service Unit, Weizmann Institute of Science, Rehovot 76100, Israel

**ABSTRACT:** Hybrid ferromagnetic/superconducting systems are well known for hosting intriguing phenomena such as emergent triplet superconductivity at their interfaces and the appearance of in-gap, spin polarized Yu-Shiba-Rusinov (YSR) states bound to magnetic impurities on a superconducting surface. In this work we demonstrate that similar phenomena can be induced on a surface of a conventional superconductor by chemisorbing non-magnetic chiral molecules. Conductance spectra measured on $NbSe_2$ flakes over which chiral alpha helix polyalanine molecules were adsorbed, exhibit, in some cases, in-gap states nearly symmetrically positioned around zero bias that shift with magnetic field, akin to YSR states, as corroborated by theoretical simulations. Other samples show evidence for a collective phenomenon of hybridized YSR-like states giving rise to unconventional, possibly triplet superconductivity, manifested in the conductance spectra by the appearance of a zero bias conductance that diminishes, but does not split, with magnetic field. The transition between these two scenarios appears to be governed by the density of adsorbed molecules.

It is well established by now, both theoretically[1-4] and experimentally[5-7] that equal-spin triplet superconductivity can emerge on the superconducting side of an interface between a conventional singlet superconductor and a ferromagnetic layer. This proximity-induced triplet order parameter has an overall symmetry that respects fermionic anti-symmetrization, e.g., a combination of odd-frequency s-wave with even frequency p-wave.[4] A related system, comprising discrete magnetic impurities (rather than a uniform magnetic layer) on the surface of a conventional s-wave superconductor attracted considerable interest in recent years. By breaking time reversal symmetry, magnetic impurities form discrete low energy spin-polarized bound states within the superconducting gap, known as *Yu-Shiba-Rusinov* (YSR) states.[8-10] The induced state is localized around an isolated impurity, and decays with an oscillating sub-structure[10] over a length scale ranging from several angstroms up to tens of nanometers, depending on the superconductor properties and dimensionality. In two- or quasi two-dimensional systems, such as $NbSe_2$, the spatial extent of the induced YSR states increases, and thus they interact stronger than in three-dimensional systems.[11] When the magnetic impurities are placed in close proximity to one another, the YSR states hybridize to form bonding and antibonding combinations. These manifest themselves in the tunneling *dI/dV vs. V* spectra as splitting of each one of the peaks in the case of parallel impurity spins, or shifting them in opposite directions in the case of antiparallel spins, with respect to the isolated impurity case.[12] Increasing the number and density of magnetic impurities, in combination with strong spin orbit coupling of the superconductor and ferromagnetic interaction between the magnetic impurities, was shown to induce topological superconductivity in various geometries such as a Fe atomic chain on a Pb surface[13] or a two-dimensional Pb monolayer grown over Co-Si islands.[14] A key signature of such topological superconductivity is the emergence of Majorana bound states; as zero-modes at the ends of a one-dimensional chain, or as chiral modes at edge of a two-dimensional island of magnetic impurites.[13,14] A recent theoretical work[15] suggested that a two-dimensional array of magnetic impurities ordered on an s-wave superconductor can also yield the much sought-for chiral p-wave superconducting symmetry. In both cases above, ferromagnetic ordering is necessary for the magnetic impurities to induce chiral p-wave symmetry. A similar system of magnetic impurities, but lacking magnetic ordering is described as Shiba glass[16] where even at the presence of a no spatial order, a finite net out-of-plane magnetization may induce topological superconductivity and hence circulating edge modes around the random dopants.

The nucleation of YSR states was achieved so far by either utilizing magnetic atoms or magnetic molecules containing one or more magnetic atoms.[11,12] A different class of molecules that were found to exhibit magnetic-like properties, but contain no magnetic atoms, is that of organic helical chiral molecules. Such molecules act as spin filters via the chiral induced spin selectivity (CISS) effect.[17,18] This effect has been employed by us to spin-polarize ferromagnets by passing current through chiral molecules.[19,20] Moreover, ferromagnetic layers were shown to be uniformly magnetized by merely adsorbing chiral molecules on their surface.[21] We have also shown that chiral molecules can, in effect, spin-polarize also the surface of a singlet s-wave superconductor, inducing there a triplet component.[22–24] In these works, scanning tunneling microscopy and spectroscopy (STM and STS) measurements were performed, showing that the superconducting order parameter in Nb (a conventional singlet *s*-wave superconductor) is altered upon the adsorption of chiral helical molecules - alpha-helix polyalanine.[22] After molecules adsorption, the tunneling spectra displayed, in general, a zero bias conductance peak (ZBCP) embedded in a gap, and their overall shape conformed to a combination of *s*-wave and equal-spin triplet chiral *p*-wave (or triplet odd-frequency s-wave) pairing potentials. A similar behavior was observed also for proximity-induced superconductivity in a gold film coupled to NbN (a conventional superconductor).[23] In these STS works the molecules were densely adsorbed over wide regions over the surface, making it impossible to differentiate between the effects of tunneling through a molecule or directly to the substrate (Nb, Au, or $NbSe_2$) next to it. Consequently, we were not able to determine from these experiments, nor from measurements performed on Nb/Chiral-molecules/graphene/Au devices,[24] whether the appearance of the in-gap states was a result of spin-polarized current generated by the CISS effect when passing through a molecule, or whether it reflected order-parameter modification at the superconductor's surface due to the chemical bonding of the molecules to it. In addition, the fact that no isolated molecules were found prevented monitoring possible effects of individual (or coupled) magnetic-like impurities and the formation of YSR states.

Following the considerations above, we set out to investigate the origin and nature of the order parameter change in conventional singlet s-wave superconductors upon chiral molecules adsorption and the possibility of YSR-like states. Particularly, we aim at answering the following question: do magnetized-like states emerge on the superconductor surface upon the adsorption of chiral molecules akin to the case of hybrid ferromagnet/superconductor systems? Namely, observing YSR-like states for low-density molecular adsorption turning into an equal-spin triplet superconducting state for high adsorbate density. To that end we expanded our previous STS-based studies and performed temperature and magnetic field dependent conductance spectroscopy measurements on devices comprising exfoliated $NbSe_2$ flakes with and without chiral alpha helix polyalanine molecules. To set the ground, however, we first applied STS to check whether the chiral-induced order parameter effect holds also for $NbSe_2$ flakes. We wish to note here that although being a singlet s-wave superconductor, $NbSe_2$ (a Van der Waals material) is more intricate than the previous systems we studied, being anisotropic and having a two-gap structure. Tunneling spectra acquired with a home-built cryogenic STM operating in a clean He exchange gas environment at 4.2K are presented in Fig. 1. On the pristine sample (Fig. 1(a)) the $dI/dV$ $vs.$ $V$ tunneling spectra conform to conventional s-wave superconductivity, while after chiral molecules adsorption (Fig. 1(b)) a clear ZBCP appeared inside the gap. This is consistent with our previous results, described in refs. [22,23] and discussed above, suggesting that the introduction of chiral molecules modifies the order-parameter.

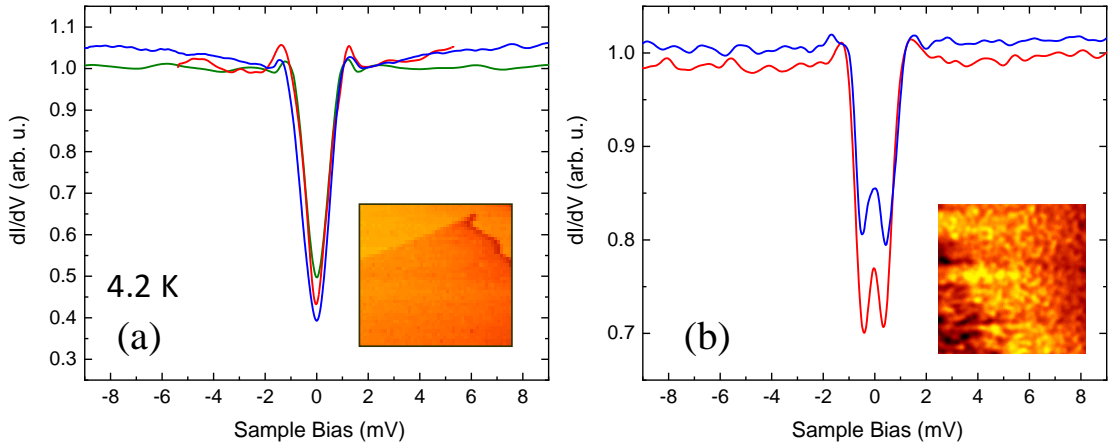

**Fig 1**: Tunneling spectra at 4.2K acquired using STM on the surface of a pre-cleaved bulk NbSe₂ crystal before (a) and after (b) alpha helix polyalanine molecules adsorption, showing conventional s-wave spectra before adsorption and zero bias conductance peak after adsorption, signifying the emergence of unconventional superconductivity. The inset of (a) shows an 800×800 nm² STM topographic image of the surface before adsorption, revealing a NbSe₂ monolayer step 6-7 A high, while the inset of (b) portrays a 60×60 nm² STM image of the surface after adsorption, revealing a non-homogenous configuration of adsorbed molecules, akin to that found in Refs.[22,23]. The spectra presented here were all measured, correspondingly, on the regions presented by the images.

The next step was the realization of device-based spectroscopy of chiral molecules adsorbed on NbSe₂. Here, two types of devices, or junctions, were examined. In the first geometry, NbSe₂ flakes were placed directly over pre-patterned Au electrodes (see Fig. 2(a)), whereas in the second (Fig. 2(d)) the flakes were positioned over a monolayer of alpha-helix polyalanine molecules chemically bonded to the Au electrodes, *but not to the NbSe₂ flake*. Importantly, in the latter (NbSe₂-flake/chiral-molecules/Au) type of devices the current must pass through the chiral molecules during measurements, and thus the corresponding role on the emergence of in-gap states can be tested. The conductance spectra measured on these devices were compared with those obtained after depositing the chiral molecules on the free surface of the NbSe₂ flakes, to which they were chemically bonded. We have found that for both geometries in-gap states emerge *only after* adsorbing the chiral molecules on top of the flakes, indicating that the CISS effect on current passing through the chiral molecules is not responsible for their appearance but rather the chemical bonding of the helical molecules to the NbSe₂ surface. As shown below, two types of states were observed, either a single ZBCP or a group of peaks positioned nearly symmetrically around zero bias. The dependence on magnetic field suggests that the former can be associated with unconventional triplet order parameter (e.g.,[25] chiral p-wave), whereas the latter with molecule-induced magnetic YSR-like states, consistent with our model simulations. The transition between these two regimes may be governed by the density and/or orientation uniformity of the adsorbed molecules, akin to the case of magnetic impurities discussed in Ref [15,16].

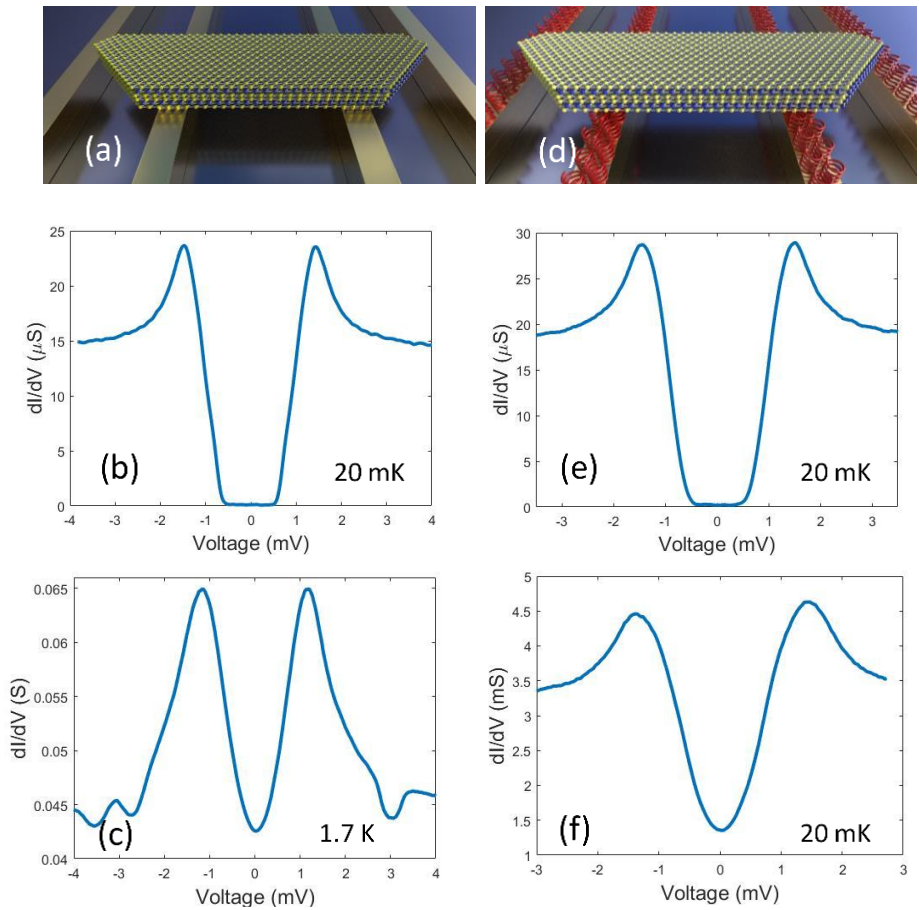

**Fig 2**: Schemes of the two types NbSe2/Au junctions studied and corresponding spectra. (a) NbSe₂ flake directly exfoliated on clean Au electrodes. (b,c) Typical *dI/dV vs. V* spectra measured on junctions of the type presented in (a), ranging from showing a hard gap (b) to the Andreev spectroscopy regime (c). (d) NbSe₂ flake exfoliated over a monolayer of alpha helix polyalanine molecules chemisorbed on Au electrodes. (e-f) Typical *dI/dV vs. V* spectra acquired from junctions of type the presented in (d), ranging from showing a hard gap (e) to a soft gap (f). Note that ZBCPs were not observed on both types of junctions.

50nm-Au/5nm-Ti thick gold electrodes were fabricated on a 300nm-SiO$_2$/Si wafer using standard procedure of direct laser-write patterning, evaporation and lift-off. Their width and center-to-center separation was 1.2µm and 5µm, respectively (see, e.g., Fig. 3(b)). The electrodes were cleaned with O$_2$ plasma, and 5-40 nm thick flakes were exfoliated from a bulk NbSe$_2$ crystal (T$_c$ ≈ 7.2K) and transferred inside a glove box either directly over the electrodes (NbSe$_2$/Au junction), or over a monolayer of alpha helix polyalanine chiral molecules chemically-adsorbed on the Au surface (NbSe$_2$/ChMol/Au junction). The monolayer was formed by inserting the electrodes into a solution of 1mM polyalanine alpha-helix molecules dissolved in ethanol for 72 hours while stirring the solution to increase uniformity. Thiolated molecules are generally known to chemically engage Au – indeed, the same method was used by us and by others extensively in prior publications.[26–32] Most junctions were measured in a dilution refrigerator with base temperature of 20-50 mK, but some in a pumped LHe system, going down to 1.5 K. Importantly, the general effect of chiral molecules adsorption on the conductance spectra was independent on flake thickness in the range studied. This may be partly due to the expected bonding of chiral molecules also to the flake edges (see below).

We have first characterized the electrical properties of the two types of junctions. The normal resistance of the NbSe$_2$/Au junctions largely varied, most of them were in the range 1-250 Ω, but some showed higher resistance up to 70 kΩ (see Supporting Information, Fig. S1, for detailed statistics of the junctions' resistances). The high resistance junctions were found in cases of bad flake-electrode contact, such as that shown in Fig. S2. Correspondingly, considerable sample-to-sample variations were found also for the *dI/dV vs. V* conductance spectra (measured using 4-probe configuration), ranging from tunneling spectra exhibiting hard gaps (Fig. 2b) to the Andreev spectroscopy regime (Fig. 2c). The NbSe$_2$/ChMol/Au junctions exhibited a smaller range of spectral behaviors, with the high resistive junctions (mostly > 5 kΩ) displaying a hard gap (Fig 2 (e)) and the low resistance junctions (300 Ω) showing a soft gap (Fig. 2(f)), but Andreev-like spectra were not measured on such junctions. It is important to note here that in-gap states for the NbSe2/ChMol/Au samples were never observed, suggesting that they are not due to current passing through the chiral molecules.

After characterizing the junctions as shown in Fig. 2, polyalanine alpha-helix molecules were deposited on the NbSe$_2$ flakes by drop-casting a solution of ~5 mM in ethanol, taking care not to modify the NbSe$_2$/Au or NbSe2/ChMol/Au junction properties. A thin non-homogeneous oxide layer was formed on the upper layer of the NbSe$_2$ flakes, as observed via XPS (X-ray Photoemission Spectroscopy), STEM (Scanning Transmission Electron Microscopy) and EDS (Energy-Dispersive X-ray Spectroscopy) measurements (for further information see Figs. S2-S5 and related discussions in the Supporting Information). Such a layer allows chemical bonding between the molecules and the NbSe$_2$ flakes, as discussed in Refs.[33,34] for similar Van der Waals materials. In particular, surface Nb-Oxide, as found in our XPS measurements, was shown to enable bonding of various organic molecules to a Nb film.[31] Furthermore, the XPS measurements corroborated the presence of a molecular monolayer on the sample even after repeated rinsing, pointing to their strong bonding to the surface. The molecules can also chemically bond to the flake edges owing to the presence of dangling bonds there,[33] but this could not have been observed in our STEM measurements. In some cases, mechanical flake deformations caused by capillary forces did alter the junction's resistance by a factor of up to two, thus limiting exact quantitative comparison between pre- and post-adsorption spectra.

We first describe the effects of chiral molecules adsorption on NbSe$_2$/Au junctions, starting with a low normal-resistance junction, $R_N$ = 120 Ω (see Fig. 3). Figure 3(a) presents temperature-dependent $dI/dV$ *vs. V* conductance spectra measured on a ~25 nm thick NbSe$_2$ flake presented in the inset of Fig. 3(b) using current-biased four probe configuration. A clear ZBCP situated within a gap in a dome-like structure appears in the tunneling spectra below 5K, persisting down to the lowest (nominal) temperature measured, 50 mK. Below 1.3 K the spectra did not change anymore, hence we do not present data acquired below 1.3 K. In Fig. 3(b) the temperature dependence of the zero bias conductance is depicted. At $T_C$ (≈ 7.2 K) the conductance at zero bias starts increasing as the temperature is lowered due to the Andreev reflection process, and the inflection point at 5.5K marks the formation of the ZBCP. We note in passing that the dips outside the gap originate from the critical current of the Au/NbSe$_2$ junction, and appear as a consequence of the low junction resistance, and their temperature dependence is tied to that of the critical current.[35]

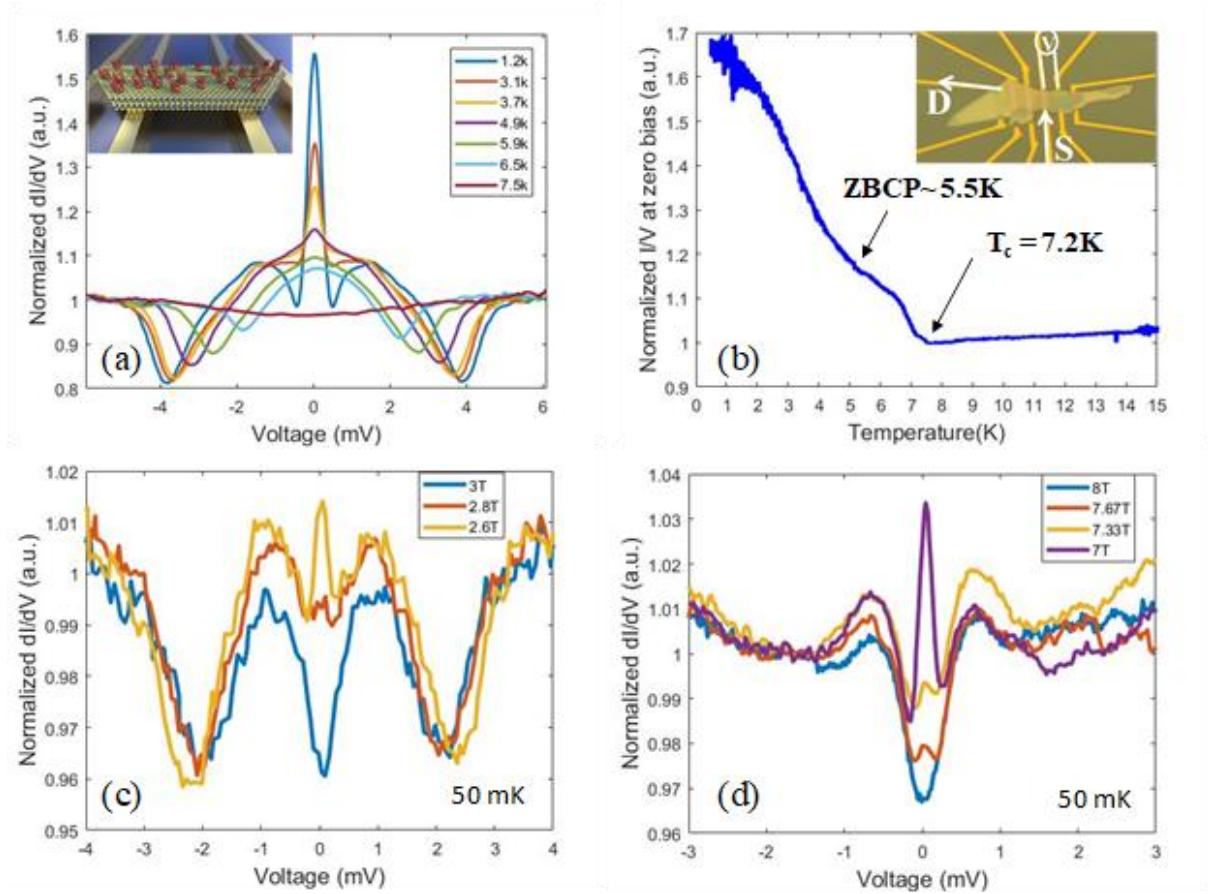

**Fig 3:** Conductance properties of a low resistance (120 Ω) NbSe$_2$/Au junction after chemisorption of chiral molecules on the NbSe$_2$ flake (~25 nm thick). (a) Temperature dependence of *dI/dV vs. V* spectra showing a distinct ZBCP that vanishes at high temperatures, yet still below Tc. Inset: Illustration of chiral-a molecules/NbSe*2*-flake/Au sample. (b) Temperature dependence of the conductance at zero bias with two transition temperatures marked by arrows: T$_C$ = 7.2 K, where the zero bias conductance starts to rise due to the Andreev dome and 5.5 K where a ZBCP starts to appear. Inset: Optical image of the sample with the measurement scheme depicted. (c-d) perpendicular (c) and parallel (d) magnetic field dependence of the conductance spectrum, showing that in high magnetic fields, yet below the parallel and perpendicular critical fields (Hc$_2$) of bulk NbSe$_2$, the ZBCP vanishes, reveling an underlying gap.

Figures 3(c) and 3(d) show the dependence of the spectra on magnetic field applied perpendicular (c) and parallel (d) to the NbSe$_2$ flake (for larger field ranges see Fig. S7). Evidently, the ZBCP vanished at perpendicular (parallel) fields of 3 T (8 T), revealing an underlying gap in the conductance spectra. Notably, the ZBCP always vanished at temperatures and magnetic fields lower than T$_C$ and the corresponding critical magnetic fields (out-of-plane, 5 T, and in-plane, 15 T), suggesting that it is associated with a sub-dominant order parameter.

Another measurement performed on a low-resistance (25 $\Omega$) junction is presented in the Supporting Information, Fig. S8. Prior to adsorption, the spectrum conforms to an s-wave superconductor/normal-metal contact in the Andreev regime, and could be fitted using the BTK formalism[36] with the a dimensionless barrier-height parameter Z=0.59. After molecular deposition, a small but clear ZBCP has emerged situated within the original gap. As in some other cases (e.g., Figs. 4,5, S7 and S8 ), the *dI/dV vs. V* spectrum became asymmetric upon adsorption, an effect that is consistent with the presence of magnetic impurities creating asymmetry between the electron and hole excitations, as discussed in Refs. [37–39].

A markedly different effect of chiral molecules adsorption is shown in Fig. 4, presenting conductance spectra measured on a ~5 nm thick NbSe$_2$ flake with high junction resistance (R$_N$=20 k$\Omega$). Due to the high junction resistance and the existence of Ohmic contacts on the same flake, two probe, voltage-biased, measurements were conducted.  The tunneling spectra before adsorption exhibit a two-band hard gap structure (Fig. 4a), consistent with the s-wave two-band density of states of NbSe$_2$[40]. After molecule adsorption, six in-gap peaks appeared, situated nearly symmetrically around zero bias, but showing a strong asymmetry in magnitude. Such asymmetry also holds for the gap edges and supra band-gap regions (see extended bias-range spectrum in Fig. S9), akin to the aforementioned effect of magnetic impurities. It should be noted here that such asymmetry *was not* seen when a single, non-split ZBCP was observed (Fig. 3), indicating that we encounter here two different phenomena, both related to the magnetic-like nature of the adsorbed molecules as discussed below. One is related to the emergence of unconventional order parameter symmetry (as reported in Refs. [22–24]), and the other for the formation of YSR states or bands. Note that in a disordered lattice of individual YSR states separated by a distance on the order of coherence length, extended quasi-particle states (or bands) can be formed, bridging the two leads. Further evidence for chiral molecules acting as magnetic-like impurities and inducing YSR bands is provided by the dependence of their position on parallel magnetic field, presented Fig. 4(c), showing a clear Zeeman-like evolution with field. The energetic shifts can be fitted with g-factor of 2 ± 0.1, with one exception – the band corresponding to the lowest bias shown in Fig. 4(c).

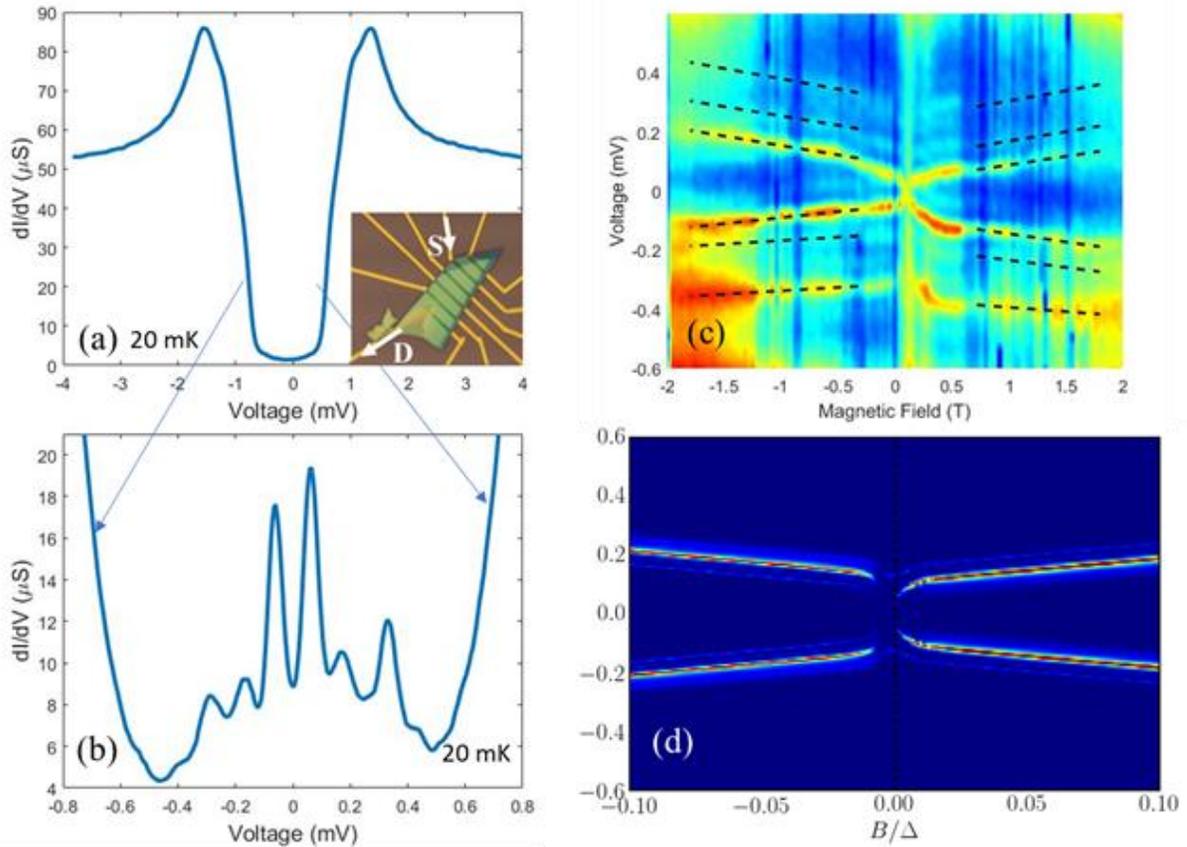

**Fig 4:** Conductance properties of a high resistance (30 kΩ) NbSe₂/Au junction. (a) *dI/dV vs. V* conductance spectrum measured before chiral molecules adsorption, showing a two-band gap structure typical of NbSe₂. Inset: Optical image of the sample showing the ~5 nm thick NbSe₂ flake with the 2-probe measurement scheme. (b) *dI/dV vs. V* spectrum measured on the same junction after molecules adsorption, showing six peaks symmetrically positioned around zero bias with and overall asymmetry in intensity. (c) Parallel magnetic field dependence of the tunneling spectra showing the divergence of the peaks position with applied magnetic field (black lines indicate peaks position). The field was swept from -2 to +2 T. (d) A simulation of the tunneling conductance of a superconductor/insulator/normal-metal junction with an array of YSR states scattered on the S side. YSR bands form and respond to the applied magnetic field in a similar way to the measurement in (c).

An important characteristic of the magnetic response in this sample is its hysteretic nature. When changing the field from -2T to 2T, the isolated conductance peaks converge at 0.2T (as seen in Fig. 4(c)), but when reversing the field scanning direction, ramping now from 2T to -2T, the picture is reversed, and the convergence point is at negative bias (not shown here). This hysteresis may be related to the polarization of the defect states generated by the chiral molecules, which is susceptible to an external magnetic field, and may shift magnetic-like domain walls. The application of the field cannot result in a full spin rotation as it is impeded by the chirality and the asymmetric structure of the molecule. This issue, along with the model simulation altogether, will be further discussed in detail elsewhere. Importantly, in a control experiment we have performed using was thioglycolic acid molecules, which should have the same binding chemistry to the $NbSe_2$ flake as that the chiral polyalanine ones, the spectra showed no signatures of unconventional superconductivity or of YSR states (see Fig. S10).

We now turn to address the $NbSe_2$/ChMol/Au junctions. As expected, these junctions exhibited, in general, much higher normal resistances compared to the "direct" $NbSe_2$/Au ones, ranging between a few hundred Ohms to G$\Omega$s and more. Recalling that before adsorbing chiral molecules on top of the $NbSe_2$ flake, only (hard or soft) gaps were observed, free of in-gap states. This stands in contrast to the spectra measured following adsorption. Here, about 10% of the junctions exhibit clear in-gap states, such as the one portrayed in Fig. 5, measured on a ~40 nm flake. At zero magnetic field, we found an asymmetric peak, which appears to be more like an unresolved double peak near zero bias, and situated within an asymmetric gap structure. Applying magnetic field parallel to the sample caused the peak to split (Fig. 5(b)) and the two resulting peaks separated monotonically with applied field conforming to g = 6. In this case, the two peaks converge at zero field and the measured spectrum did not depend on the ramp direction of the magnetic field, unlike the behavior observed in Fig. 4(c).

One can draw the following conclusions from our experimental results delineated above. First, that chemical bonding of the chiral alpha helix polyalanine molecules to the $NbSe_2$ surface is required for the in-gap states to emerge, and not the mere passing current through the molecules. Second, that these molecules appear to act as magnetic-like impurities when adsorbed on the surface, giving rise to either YSR-like states or emergence of unconventional order parameter symmetry consistent with triplet-pairing superconductivity. In the following, we suggest a possible mechanism by which chiral-helical molecules can act as magnetic impurities on the NbSe₂ surface upon chemisorption, and also an explanation for the transition from the discrete YSR-like state to the unconventional, triplet-pairing, superconductivity regimes.

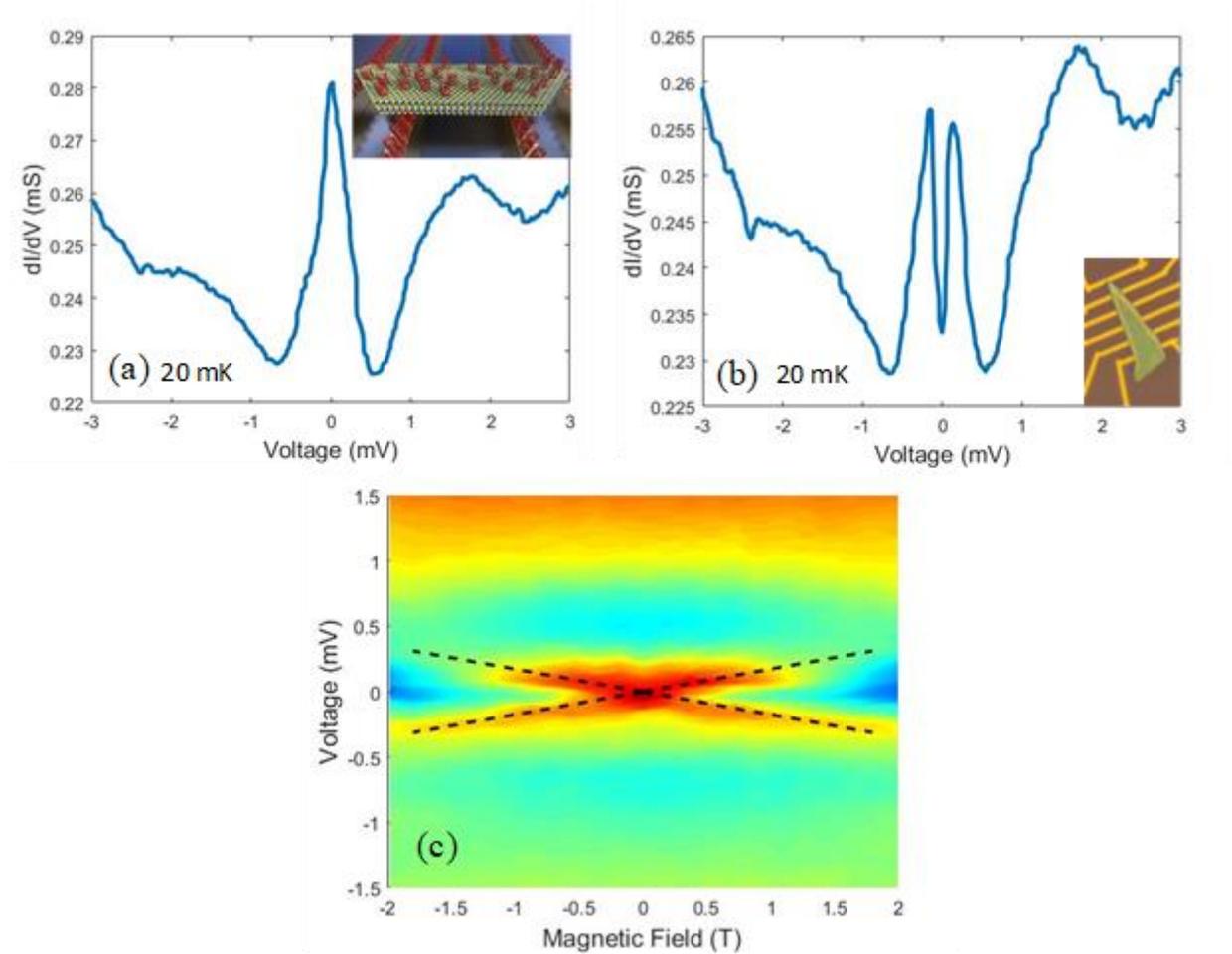

**Fig 5:** A NbSe2/ChMol/Au junction that showed no sub-gap features, but after a second step of poyalanine molecules deposition, this time *over* the NbSe₂ flake (~30 nm thick), exhibited a pronounced ZBCP. (a) *dI/dV vs. V* spectrum after second molecules adsorption and device's illustration (inset). (b) Tunneling spectrum under 1 T magnetic field applied parallel tom the flake, exhibiting peak splitting. (inset) optical image of the device prior to the second adsorption. (c) Parallel field dependence of the tunneling spectra showing the divergence of the split peaks position with applied positive to negative magnetic field (black lines indicate peaks position).

So far, theoretical and experimental works showed that YSR states can be induced by magnetic adatoms or magnetic molecules (molecules containing one or more magnetic atoms) physisorbed on s-wave superconductors.[11,12,41] In contrast, the alpha-helix polyalanine molecules we study here do not contain a magnetic atom, and unlike previous works with magnetic molecules,[12] chemical bonding is needed to induce YSR-like behavior. The adsorption of alpha-helix polyalanine molecules has been shown to induce a variety of magnetic phenomena such as uniformly magnetizing gold-coated Ni films,[21] inducing soft ferromagnetism in gold[42] and turning super-paramagnetic nanoparticles ferromagnetic.[43] Moreover, coexistence of superconductivity and ferromagnetism was detected in NbSe$_2$ flakes upon the adsorption of polar reductive hydrazine molecules due to surface-structural modulation,[44] deeming the possibility of magnetizing NbSe$_2$ via molecular adsorption possible. However, as for yet, there is no theoretical model describing the formation of YSR states due to adsorbed chiral molecules.

The conduction response to applied magnetic field shown in Figs. 4(c) and 5(c) is also consistent with the YSR-like states picture and can be attributed to the formation of YSR bands, constructed from an array of coupled magnetic impurities on a superconducting surface. Furthermore, multiple (e.g., six) peaks may result from different bonding configurations, consistent with the observation of Choi et al.[41] that different orbitals in a magnetic atom adsorbed on a superconductor may result in distinct sub-gap energies. The formation of the bands and part of their response to the magnetic field can be explained via our simulation of the system (see Fig. 4(d) and Supporting Information for further details). In the simulation we solved coupled Bogoliubov-de Gennes equations on a two-dimensional lattice with a homogeneous order parameter and spin-orbit interaction. Classical magnetic impurities were distributed with a separation of a few lattice constants. Differential conductance of the sample was derived from the S-matrix which describes the scattering between two normal leads, assuming a finite life-time for the quasiparticle excitations. The form of the bands in Fig. 4(c) suggests that the polarization of the impurities has a non-trivial dependence on the magnetic field. To account for that behavior, we assumed that the spin of the impurity has a linear dependence on the magnetic field for small values of the field. For higher values, however, we assumed a saturation of the response due to the complete alignment of the impurity spins with the field. We note in passing that our simulations treated an isotropic, single band BCS superconductor, unlike NbSe$_2$, but we believe that it does capture the main physics underlying the emergence of YSR-like bands behavior.

The results in Fig. 3 pose a greater challenge since only one peak at zero energy exists and an applied magnetic field suppresses the ZBCP but does not cause any splitting associated with YSR states. One possible trivial explanation for the existence of pronounced zero bias peak is a very small critical junction current manifested as a ZBCP. While such an explanation is possible, it is important to note that critical current is strongly temperature dependent around Tc (~ 7.2 K in our case) and below 0.5Tc should not show strong variations with temperature. In Fig. 3(a), however, the magnitude of the ZBCP is seen to diminish in the range of 1.3K to 5K, while hardly notable above 5K. More so, for other samples with lower resistance, where larger currents had to be supplied in order to probe the voltage comparable to Δ (1.2 meV), no such peaks were found. A different possible scenario that can account for the results in Fig. 3, as well as previous results measured using STM, is the emergence of a topological superconducting phase, mainly chiral p-wave order parameter symmetry. Most models for such topological phases demand a uniform ferromagnetic ordering; a condition that probably does not occur in our system considering the drop cast method we used to form the molecular monolayers. Indeed, our previous STM/STS studies[22,23] of chiral molecules adsorbed on Nb and Au/NbN films showed that the adsorption is not uniform and homogeneous all over the surface, although we did find islands of uniformly aligned molecules within an overall disordered molecular layer (including molecule-free regions). A similar disordered adsorption configuration was revealed also in our present STM measurements on the surface of a cleaved NbSe$_2$ crystal (e.g., inset of Fig. 1(b)). This non-uniformity may result for the non-homogeneous oxide layer evidenced in our XPS measurements. We would like to emphasize that even under these adsorption conditions the helical chiral alpha-helix polyalanine molecules used here were found able to polarize ferromagentic layers and modify the order-parameter symmetry of s-wave superconductors. The axis of the alpha-helix molecule breaks the spherical spin symmetry and induces a favorable direction for spin alignment at the instant of adsorption. Therefore, the orientation of each molecule is likely to be closely related to the direction of the induced magnetic moment upon adsorption. Hence, we believe that our system is similar to incorporating non-ordered magnetic moments on the superconductor surface. This system is described as a Shiba glass,[16] where even at the absence spatial order, a finite net out-of-plane magnetization could induce topological superconductivity and hence circulating edge modes around the random dopants. Nakosai et al. treated theoretically a system of magnetic moments placed on the surface of an s-wave superconductor, inducing unconventional $p_x + p_y$ or chiral $p_x + ip_y$ conductivity, depending on the spin orientation configuration of these moments.[45] Unfortunately, we are not able to control the orientation of the adsorbed alpha helix molecules, obviously not to the degree described in the paper, and therefore we are not able to test the detailed predictions of the authors.

One question that still remains is why we observe in some cases YSR-like states and in other cases spectra that conform to unconventional superconductivity. This may be due to the effect of chemisorbed molecular density and uniformity over the surface, probably governed by the non-homogenous surface oxidation distribution, as revealed by XPS. As shown in Ref. [16], above a critical density, the YSR states hybridize to form a Shiba glass system, and even in the presence of random spatial moment distribution a net out-of-plane polarization is formed. In this case, above some critical density topological superconductivity can be induced. This model may account for the appearance of pronounced ZBCPs in both our STM and device measurements (such as Fig.3) as the result of an emergent sub-dominant triplet p-wave superconductivity. As shown in Fig. 3, the unconventional superconducting component vanishes at temperatures and magnetic fields well below $T_C$ and $H_C$.

In summary, we measured the conductance spectra of $NbSe_2$/Au and $NbSe_2$/ChMol/Au junctions coated with a layer of alpha-helix polyalanine molecules and found in some cases that chemically adsorbed chiral molecules induce a magnetic like state behavior that is consistent with YSR states formed by magnetic impurities on a superconducting surface. Specifically, in-gap states appeared upon adsorption, nearly symmetrically positioned around zero bias, shifting to higher energies with the application of magnetic field. Model simulations showing the formation of YSR-like bands from interacting YSR states randomly dispersed on the superconducting surface qualitatively reproduced these results. In other cases, we observed the emergence of a strong robust ZBCP, which points towards the emergence of topological triplet p-wave superconductivity. In this case it is reasonable to assume that the adsorbed disorder layer acts as a Shiba glass due to the spherical symmetry breaking of the adsorbed layer while the helicity of the molecule favors a given spin.  Importantly, our present work provides further strong evidence that chiral helical molecules act, upon chemisorption to a superconductor surface as magnetic impurities, opening possibilities for their use in superconducting-spintronic[1] devices.

**ASSOCIATED CONTENT**

**Supporting information**

Additional conductance spectra, XPS, STEM and EDS results, junction resistance statistics and details of model calculation.

**AUTHOR INFORMATION**

**Corresponding Authors**

Email: milode@mail.huji.ac.il, paltiel@mail.huji.ac.il

**ACKNOWLEDGMENTS**

We thank Ting-Kuo Lee and Chao-Cheng Kaun for helpful discussions.  The research was supported in parts by the ISF F.I.R.S.T. program (grant # 687/16), the Niedersachsen Ministry of Science and Culture grant (O.M. and Y.P.), a grant from the Academia Sinica – Hebrew University Research Program (O.M., Y.P. and C.C.K.), and project number AS-iMATE-107-95 (T.K.L. and C.C.K.). O.M. thanks support from the Harry de Jur Chair in Applied Science. KY and EG acknowledge support from the Israel Science Foundation under Grant No. 1626/16. HS is supported by European Research Council Starting Grant 637298 (TUNNEL). T.D. is grateful to the Azrieli Foundation for an Azrieli Fellowship.

# Supporting information

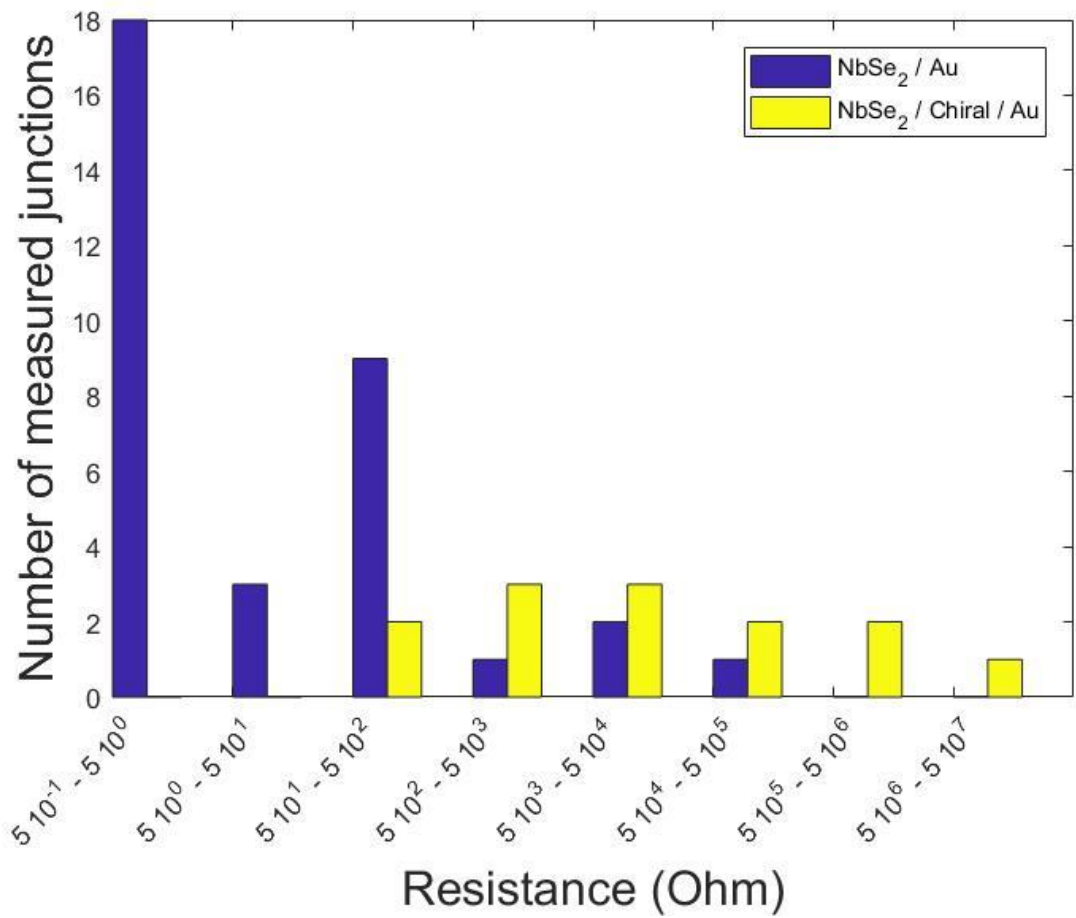

**Fig S1:** Histogram of junction resistances for the two types of junction configurations: NbSe$_2$-flake/Au (blue columns) and NbSe$_2$-flake/Chiral-Molecules/Au (yellow columns) junctions. Evidently, the presence of chiral molecules on the Au electrode tends to increase the junction resistance. The junctions of largest resistance (> 500 $\Omega$) in the NbSe$_2$-flake/Au configuration may be due to deformation of the NbSe$_2$ flake, degrading the contact, and the smallest junction resistances (< 1 k$\Omega$) in the NbSe$_2$-flake/Chiral-Molecules/Au configuration are probably due areas with very dilute molecular coverage.

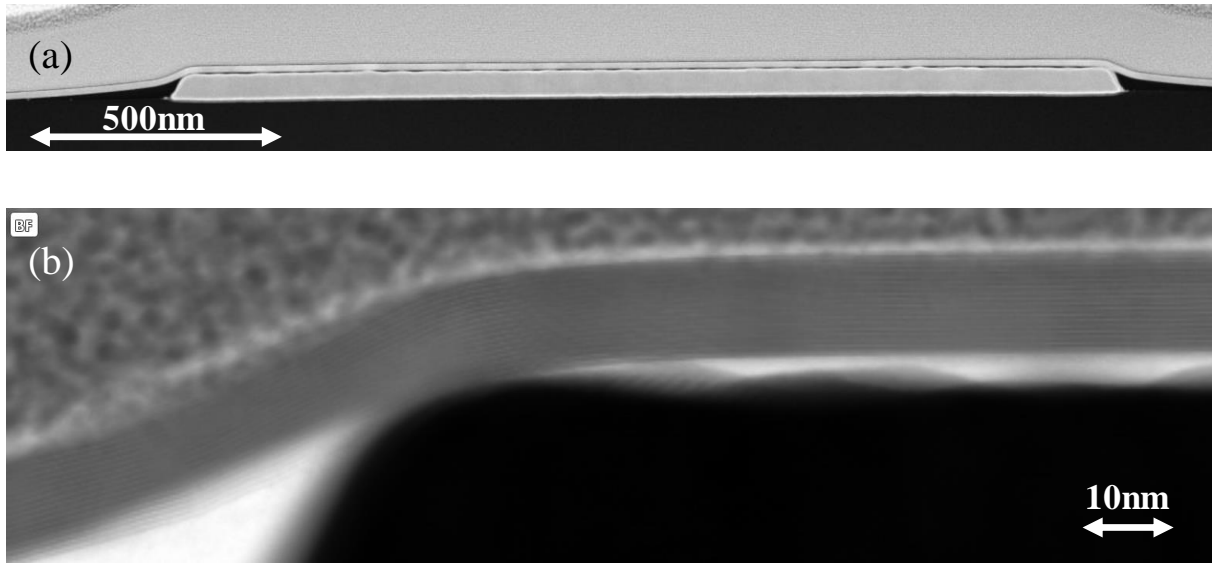

**Fig S2:** STEM images of a NbSe$_2$/Au-electrode contact cross-section, showing a poor contact between the flake and the electrode. (a) Dark field STEM image of the interface between a NbSe$_2$ flake and an entire electrode. Flake bending (which impacts the contact resistance) on the two sides of the electrode is visible. (b) Bright field STEM image focusing on the interface between the corner of an electrode and a bended NbSe$_2$ flake. Other than the corner, a poor contact with between the flake and the electrode is evident from the gaps that are clearly seen. The NbSe$_2$ flake (in both images) is covered with a thin platinum film deposited for the cross-sectional STEM measurements.

## Experimental - XPS

XPS measurements were performed on a Kratos AXIS Ultra DLD spectrometer, using a monochromatic Al kα source at low power, 15-75 W, and detection pass energies of 20-80 eV. The base pressure in the chamber was $5 \cdot 10^{-10}$ torr.

Due to differential charging artifacts encountered with these samples, controlled surface charging (CSC)[1] was applied, thus allowing the differentiation of different sample domains, as previously demonstrated in a variety of reports [see e.g. ref. [2]]. Consequently, for the presented spectra, raw energy-scale is used (not calibrated), exploiting in turn the charging shifts as a source of useful information. The electron flood gun (eFG) in these experiments was operated at 2 A filament current and 5 V grid bias.

Throughout this work results from three types of samples sample are presented: Bare NbSe$_2$ flakes on SiO$_2$ substrate, NbSe$_2$ flakes on SiO$_2$ covered molecules using two adsorption methods: drop cast and submersion in a molecules solution. The drop cast method guaranties a thick molecular layer even without any chemical bonding, while the submersion method entails repeated rinsing of the sample after molecules deposition which would leave mainly bonded molecules and so supports the formation of a molecular monolayer.

The evolution of damage under the x-ray beam was carefully inspected, starting with fast scans at low x-ray flux and, gradually, improving statistics by increased exposures. We found the inspected polyalanine molecules to be sensitive to the irradiation, showing mainly degradation in the amide C 1s peak. This effect, demonstrated in Fig. S3(c), led us to preferably rely on 'early' scans, even in cases where compromise as of the noise level was needed, while later scans still served as useful reference data.

## Results

Fig. S3(a,b) presents results taken on the drop cast sample, thus showing strong and clear signs of molecules while Fig. S3(c,d) is presenting data measured on the monolayer sample. The detection of the molecules in the drop-cast sample is much clearer since the density and thickness of the molecular layer give a pronounced signal while the monolayer sample has a much thinner molecular layer, thus yielding a weaker signal that is also affected by the molecules-substrate interaction. The presence of the molecules, however, was clearly detected in both samples.

Analysis of the C 1s and N 1s lines directly verifies the presence of polyalanine molecules and, importantly, the preservation of their chiral structure. As demonstrated in Fig. S3(a), the polyalanine C 1s line consists of three major components, indicative of the peptide structure. Interestingly, the high-energy component at ~289 eV (3.0 eV higher than the main peak) shows a clear shoulder at its high binding energy side. This shoulder is attributed to carbonyl groups that participate in hydrogen bonding to $NH_2$ end-groups of the alkane arms. The latter assignment relies on a recent work by U. Shimanovich et al. ["Towards designing material properties of protein nanofibrils via hydrogen bonds", in preparation], where the assembly of peptides at various length scales could be correlated with their efficiency to form hydrogen bonds between neighboring molecules. Here, in contrast to the latter, the hydrogen bonds are formed *within* the molecule, thus stabilizing its helical shape. Accordingly, the magnitude of the related component in the C 1s line is quantitatively very close to the theoretical value, see Table S2. This result strongly supports our conclusion that the molecules residing on the NbSe$_2$ flakes *preserve their chirality signature*. A high-binding-energy shoulder is found at the N 1s line as well, Fig. S3b,d, in agreement with the work by Shimanovich and, again, its relative intensity is quantitatively close to the theoretical value for a non-damaged molecule. Therefore, one can conclude that polyalanine binding is likely successful.

In this respect, it should be noted that only limited stability of these molecules under the x-ray irradiation is found and, most pronounced, gradual degradation was observed in the characteristic amide line, Fig. S3c, around 289 eV. The damage is expressed in finite deviations of the measured atomic concentrations, Table S2, from the theoretical values, which in fact evolved further towards prolonged exposures on any given spot of a sample. With this reservation as of the accuracy of our quantitative data, a conclusion can yet be drawn: initially, except for small amounts of contaminating hydrocarbons that were not removed by the washing steps, most of the detected signals from organic molecules correspond to non-defected polyalanine.

A second central difficulty in the XPS analysis of these samples regards the emergence of differential charging under the x-ray beam: domains that are electrically different undergo different charging magnitudes and, thus, introduce complication to the chemical analysis. To overcome this difficulty, such as to get 'the correct' chemical analysis, we performed scans at different charging conditions, following the controlled surface charging (CSC) technique.[1,2] Demonstration of peak shifts is shown in Fig. S4 for the Nb 3d and Se 3d signals. Shifts of the Si, O and even the N and C lines, are significantly simpler (not shown). Overall, the CSC analysis reveals three leading types of domains: (1) NbSe$_2$ flakes that undergo shifts by $\Delta_1$=6.3-6.6 eV; (2) domains consisting of NbSe$_2$ plus various non-homogeneous oxidized compositions, the extreme of which is Nb$_2$O$_5$, responding by $\Delta_2$=7.0-7.5 eV; (3) exposed silica domains characterized by $\Delta_3$=5.7-6.0 eV. Fig. S4 exemplifies how domains '1' and '2' can be differentiated by the CSC analysis.

While these charging effects complicate the overall analysis, an important observation arises here regarding the molecular signals: both the related C and N signals obey $\Delta_1$ shifts, indicating that their location is mainly on the NbSe$_2$ flakes. We anticipate that non-bonded molecules originally residing on (but not necessarily directly touching) open silica domains, or highly oxidized Nb, were efficiently washed out during sample preparation, leaving only the bonded molecules identified by our XPS measurements. An important message provided here regards the indirect evidence about efficient bonding of molecules to the dichalcogenide flakes. We further tried to extract direct information on the molecule-flake bonding, however encountered difficulties for the following reasons: (1) the signals from bonding sites are necessarily very weak in any realistic configuration to be considered; (2) bonding through the thiol group requires detection of weak S signals that overlap with significantly stronger Se signals; (3) due to beam damage, any of the above options could not be pushed to have the signal to noise ratio improved significantly.

**Table S1**: Raw XPS-derived atomic concentrations for the molecularly coated platelets.

|  | C | N | Nb red | Nb ox | Se red | Se ox | O | Si |
|---|---|---|---|---|---|---|---|---|
| **Atomic %** | 11.49 | 2.58 | 0.27 | 0.22 | 0.52 | 0.13 | 56.42 | 25.45 |

**Table S2**: Representative atomic concentration ratios, evaluated for signals characteristic of the amide group and the related Hydrogen bonds (labeled 'H'). Also shown is the evaluated stoichiometry at the non-oxidized domains of the $NbSe_2$.

|  | N/C(am) | 'H'/C(am) | 'H'/N | Se(r)/Nb(r) |
|---|---|---|---|---|
| **theory** | 1.194 | 0.20 | 0.17 | 2.0 |
| **Exp.** | 1.5 | 0.19 | 0.13-0.21 | 1.93 |

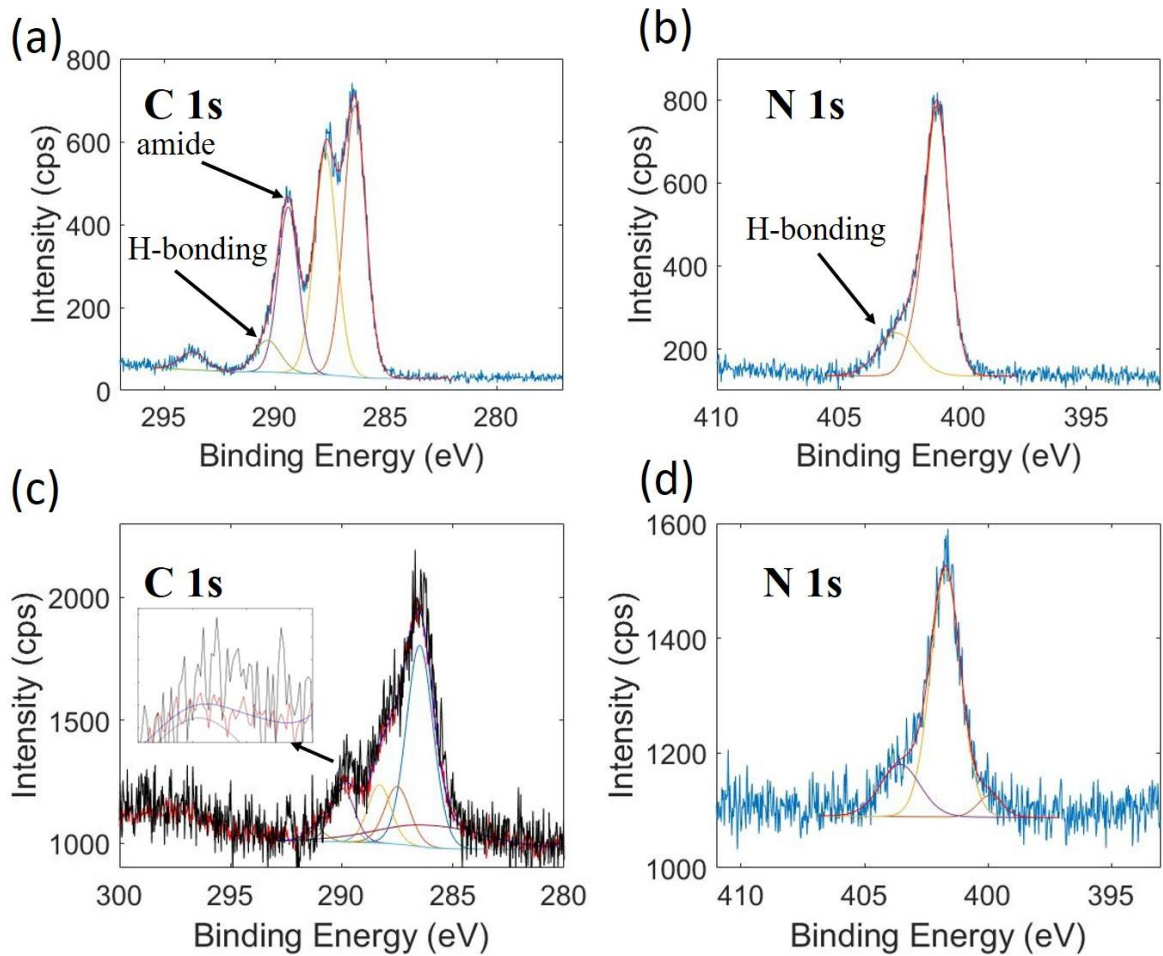

**Fig. S3**: Characteristic polyalanine C 1s and N 1s spectral windows, as recorded without charge neutralization from drop cast samples, (a) and (b), respectively, and from the molecularly coated samples, (c) and (d), respectively. Note the three major components in the carbon line of panel (a), which are characteristic to peptide systems. Note also the shoulders around 289.8 eV and 402.9 ev for the C and N, respectively, attributed to the formation hydrogen bonds. The inset to panel c shows the degradation in amide signal during exposure to the x-ray source.

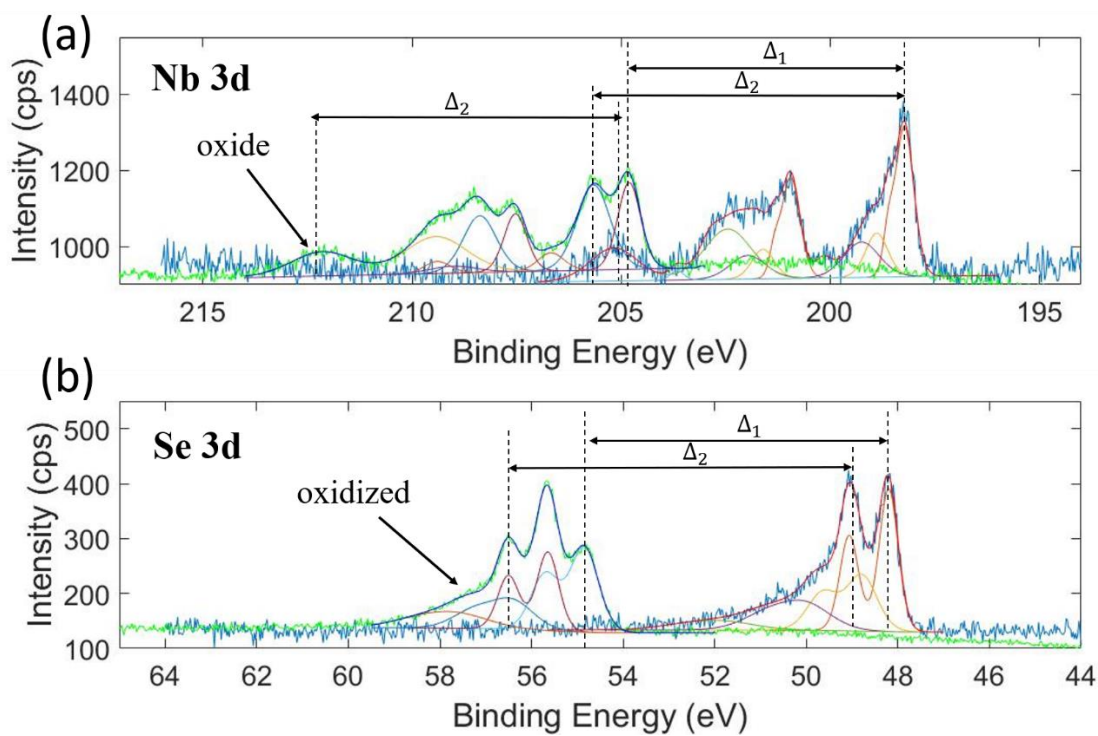

**Fig. S4**: Spectra recorded from the very same spot under two distinctly different charging conditions, eFG off and on, showing the Nb 3d spectrum (a) and the Se 3d spectrum (b). Note the intermix of charging effects with the chemical shifts and the extraction of two dominant Nb states in (a): a diselenide and an oxide state. Intermediate oxidation states do appear, yet with relatively low intensity. Corresponding analysis apply for the Se as well, but in the latter any highly oxidized Se probably evaporated prior to the XPS measurement.

**STEM and EDS measurements**

A cross-section extracted from a sample of identical configuration to the ChMol/NbSe₂/Au samples we measured (Figs. 3,4) was prepared using focused ion beam and then measured in a STEM system equipped with EDS for elemental detection and analysis. Fig. S5 shows a cross-section taken from the NbSe₂/Au-electrode interface, and Fig. S5 presents the cross-section taken from the NbSe₂/SiO₂ interface. Both figures show a 30nm thick layer of chiral alpha-helix polyalanine molecules, identified by the presence of Nitrogen (which was not present on a bare, molecule-free, sample). An oxide layer was detected on the surface of the NbSe₂ in both cross-sections.

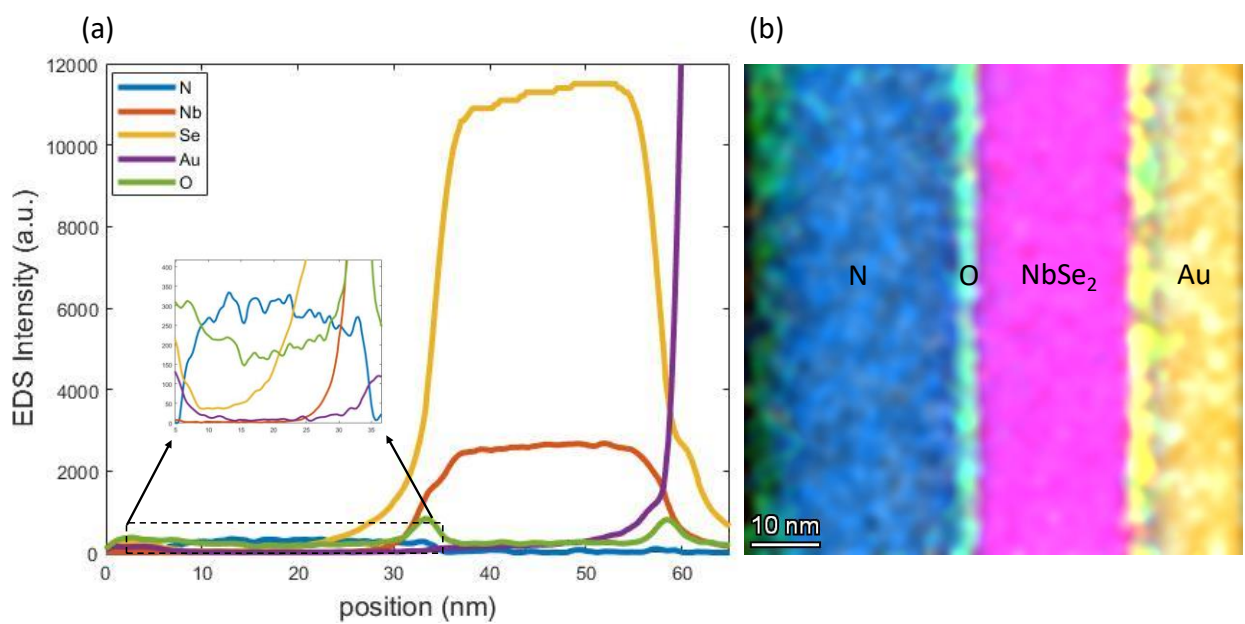

**Fig S5:** STEM-EDS measurements on a ChMol/NbSe₂/Au-electrode sample cross-sections. (a) EDS of sample cross-section showing surface oxidation of the NbSe2 flake (seen as a peak in the oxygen layer and reduction in the Nb and Se signals). The molecular layer is identified through the Nitrogen signal (prevalent in the polyalanine molecules), to the right of the NbSe2 flake, see inset.  (b) A STEM image of the sample combined with color from the EDS signal. From right to left: Au (yellow), NbSe2 (pink), oxide layer (light blue) and the molecular layer (blue). The measured molecular layer thickness is about 30nm.

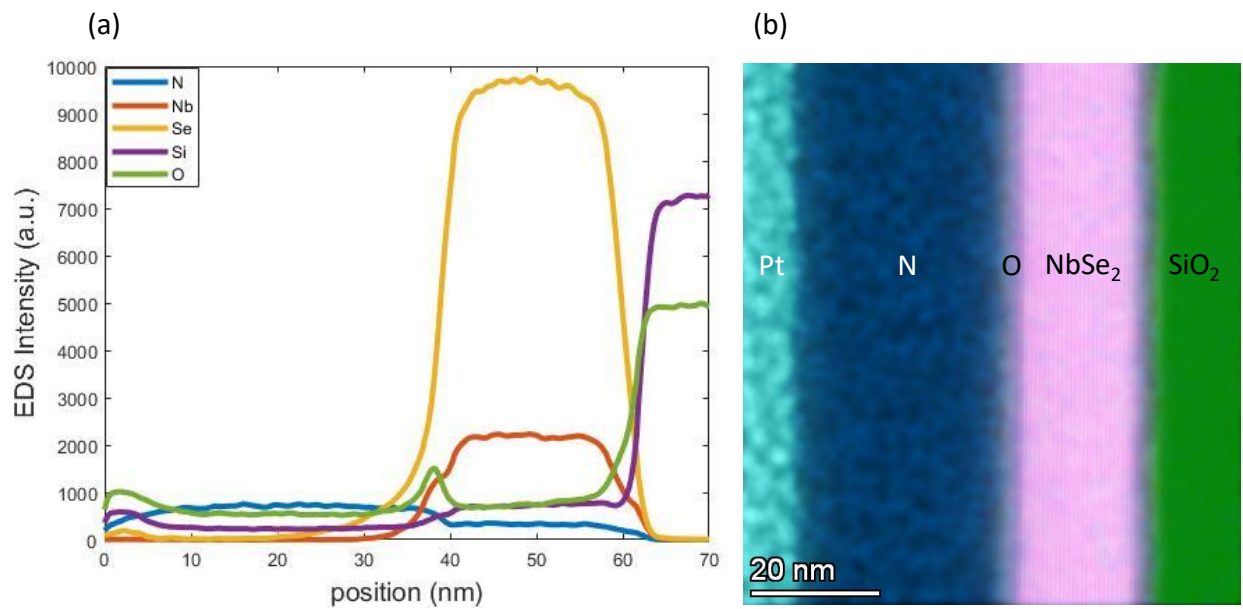

**Fig S6:** STEM-EDS measurements on a ChMol/NbSe$_2$/SiO$_2$ cross-sections. (a) EDS cross-section showing a surface oxidation of the NbSe$_2$ flake (seen as a peak in the oxygen signal and reduction in the Nb and Se signals). The molecular layer detected by the Nitrogen signal to the right of the oxidized NbSe$_2$ flake. (b) A STEM image of the sample combined with color from EDS signal. From right to left: SiO$_2$ (green), NbSe2 (pink), oxidized layer (light blue), molecular layer (blue) and Pt coating (light blue). The latter was used for the cross-section preparation. The molecular layer is about 30nm think in a typical sample.

# Additional conductance spectra

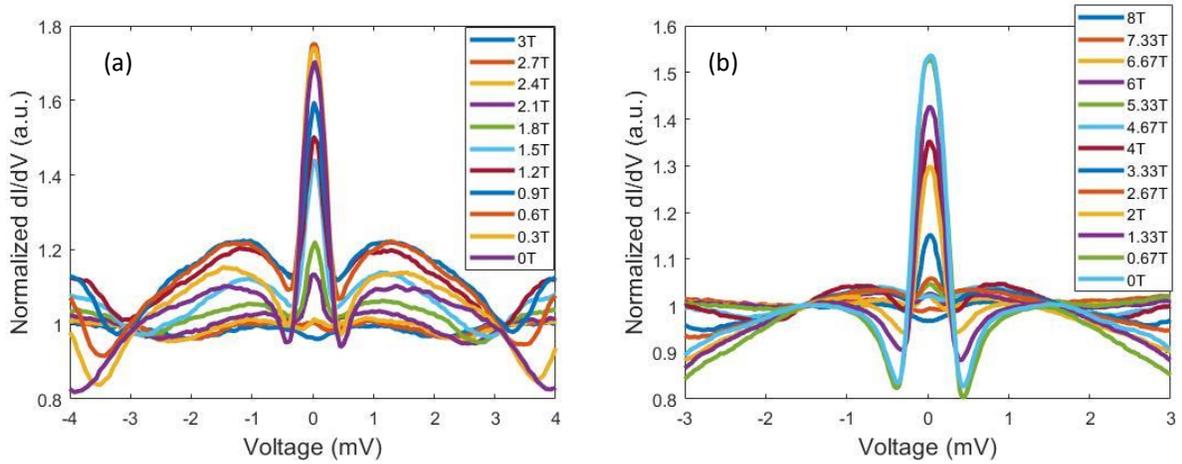

**Fig S7:** Perpendicular (a) and parallel (b) magnetic field dependence of the conductance spectra as in figures 3(c) and 3(d), respectively but over the full measured field range. Note the monotonic decrease of the peak at zero bias with applied magnetic fields.

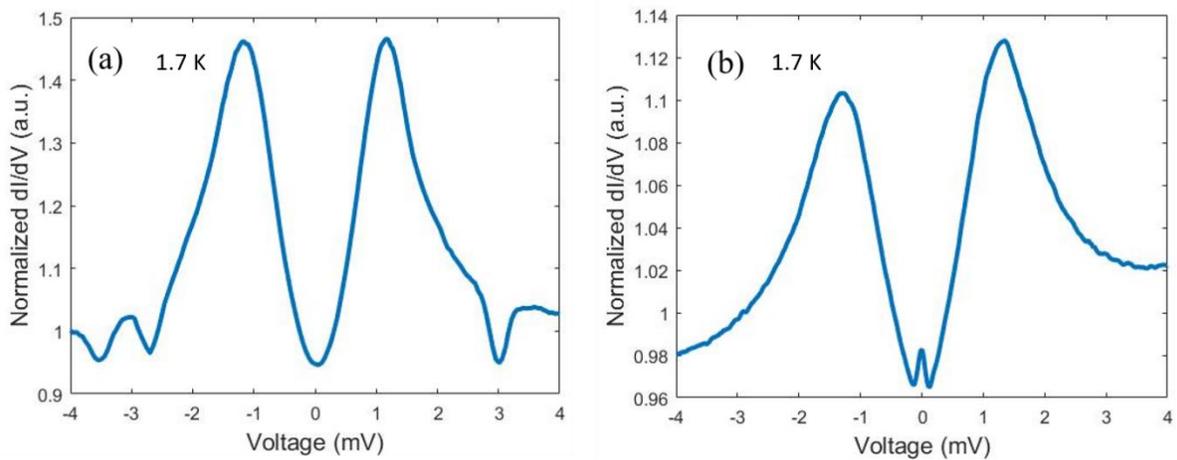

**Fig S8:** Conductance spectra measured on a $NbSe_2$-flake/Au junction in the Andreev regime (a) before adsorption, and (b) after adsorption, exhibiting a small ZBCP peak.

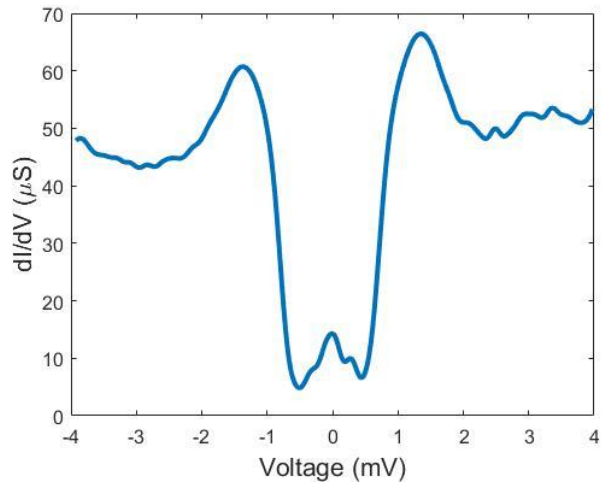

**Fig S9:** Conductance spectrum measured on the junction presented in Fig. 4(b), but over wider sample-bias range, manifesting spectral magnitude asymmetry between positive and negative bias.

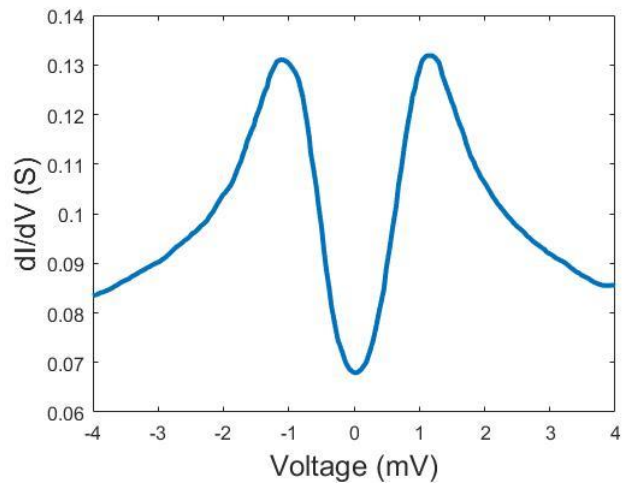

**Fig S10:** Conductance spectrum measured on a control sample with non-chiral molecules, namely, on a thioglycolic-acid-molecules/NbSe2-flake/Au-electrode junction, showing no signatures of unconventional superconductivity and no traces of YSR-like states.

## Model Hamiltonian

In this section we present the model Hamiltonian which we used for our numerical simulations and the calculation of conductance. We consider a situation where magnetic impurities are embedded in a two-dimensional superconducting lattice, which we view as a model for a superconducting NbSe$_2$ flake with adsorbed chiral molecules. The Hamiltonian is based on a mean-field tight-binding form of the Bogoliubov-de Gennes Hamiltonian, and is given by $H = H_{sc} + H_R + H_{mag}$. The term describing the superconducting part of the sample is:

$$H_{sc} = \sum_{r,\sigma} [t_x \psi_\sigma^\dagger(\boldsymbol{r} + a\hat{\boldsymbol{x}}) \psi_\sigma(\boldsymbol{r}) + t_y \psi_\sigma^\dagger(\boldsymbol{r} + a\hat{\boldsymbol{y}}) \psi_\sigma(\boldsymbol{r}) - \frac{\mu}{2} \psi_\sigma^\dagger(\boldsymbol{r}) \psi_\sigma(\boldsymbol{r}) + h.c.]$$
$$+ \sum_{r} \Delta \psi_\uparrow(\boldsymbol{r}) \psi_\downarrow(\boldsymbol{r}) + \Delta^* \psi_\downarrow^\dagger(\boldsymbol{r}) \psi_\uparrow^\dagger(\boldsymbol{r})$$

where $t_x$ and $t_y$ are the hopping amplitudes in each direction, $\mu$ is the chemical potential and $\Delta$ is the pairing strength, which we assume to be homogeneous throughout. The operators $\psi_\sigma^\dagger(\boldsymbol{r})$ and $\psi_\sigma(\boldsymbol{r})$ create and annihilate an electron with spin $\sigma = \uparrow, \downarrow$ at a discrete position $\boldsymbol{r} = ma\hat{\boldsymbol{x}} + na\hat{\boldsymbol{y}}$ on the lattice ($m, n \in \mathcal{Z}$). To account for the edge effect of the flake, we also include a Rashba spin-orbit interaction via:

$$H_R = \alpha_R \sum_{r} \Psi^\dagger(\boldsymbol{r} + a\hat{\boldsymbol{x}}) \sigma_x \Psi(\boldsymbol{r}) + \Psi^\dagger(\boldsymbol{r} + a\hat{\boldsymbol{y}}) \sigma_y \Psi(\boldsymbol{r}) + h.c.,$$

where we defined the spinor $\Psi(\boldsymbol{r}) = (\psi_\uparrow(\boldsymbol{r}), \psi_\downarrow(\boldsymbol{r}))^T$. The strength of the Rashba interaction is governed by the parameter $\alpha_R$. The coupling of the electrons to the magnetic impurities and their response to an external Zeeman field is given by:

$$H_{mag} = \sum_{r} \Psi^\dagger(\boldsymbol{r}) \left( \boldsymbol{B} + \sum_{i} J\boldsymbol{S}_i \right) \cdot \boldsymbol{\sigma} \, \Psi(\boldsymbol{r}) .$$

The magnetic impurities are located at positions $\boldsymbol{R}_i$, each having a spin $\boldsymbol{S}_i$. The impurities couple to the electrons in their vicinity via an exchange interaction with strength $J$. Similar to the usual treatment of YSR states, we assume that the impurities lack quantum dynamics and can be viewed as classical degrees of freedom.

In order to demonstrate the collective effect of the impurities, as observed in the experimental measurement of conductance in Fig. (4c), we consider impurities which are not completely polarized and thus retain a certain degree of dependence on the magnetic field. We set a population of impurities such that in the absence of a magnetic field the polarization has a preferred direction, akin to regular YSR states. However, each impurity also has a small random component, whose direction can be altered by the application of the field. Consequently, the average over all impurity-spins in the sample gives us $\langle J\boldsymbol{S}\cdot\widehat{\boldsymbol{B}}\rangle \approx J_0 + J_1 tanh\big(\boldsymbol{B}\cdot\widehat{\boldsymbol{B}}/B_0\big)$. The parameters $J_0$, $J_1$ give us the ratio between the polarized and random components, and $B_0$ is chosen to fit the experimental data. To obtain this form we assumed that the random component is free to rotate with the field and thus can achieve a complete alignment.

To obtain the conductance we look at a simple model where the two-dimensional sample is coupled to two biased leads, each via a tunnel junction. This allows us to define a unitary scattering-matrix between the two leads, which we find using the Weidenmuller formula:

$$\mathcal{S}(E) = \boldsymbol{1} - iW\left(\frac{1}{E - \mathcal{H} + i\dfrac{W^\dagger W}{2}}\right)W^\dagger ,$$

with $W$ the matrix that describes the coupling of the scatterer $\mathcal{H}$ to the leads. Here $\mathcal{H}$ is the system Hamiltonian written in a particle-hole basis. The conductance $G(V)$ is obtained by extracting the transmission and reflection amplitudes from the scattering.[3,4] To avoid the common conductance divergences at $E = \pm\Delta$ we have constrained the spectrum by adding a finite quasiparticle lifetime $\Gamma^{-1}$, with the replacement $\mathcal{H} \rightarrow \mathcal{H} - i\Gamma$.

In the numerical simulation, seen in Fig. 4(d), we used a lattice of size $55x55$ sites with impurities scattered at distances of $5 - 6$ lattice sites. The physical parameters are given by: $t_x = t_y = 1$, $\Delta = 0.5$, $\alpha_R = 0.1$, $\mu = 0.0$ and $\Gamma = 0.0025$. The parameters of the effective exchange interaction are: $J_0 = 2.1$, $J_1 = 0.3$ and $B_0 = 0.005$.